\documentclass[twocolumn,showpacs,preprintnumbers]{revtex4}
\usepackage{amssymb}
\usepackage{amsfonts}
\usepackage{amsmath}
\usepackage{graphicx}
\usepackage{dcolumn}
\usepackage{bm}

\input{tcilatex}

\begin{document}

\title{Quantum view on contextual logic of composite intelligent devices}
\author{E. D. Vol}
\email{vol@ilt.kharkov.ua}
\affiliation{B. Verkin Institute for Low Temperature Physics and Engineering of the
National Academy of Sciences of Ukraine 47, Lenin Ave., Kharkov 61103,
Ukraine.}
\date{\today }

\begin{abstract}
Based on the ideas of quantum theory of open systems (QTOS) we propose the
consistent approach to study probabilistic many-valued propositional logic
of intelligent devices that are composed from separate but interconnected
logical units.In this preliminary communication we consider only the
simplest example of such systems,namely, four- valued probabilistic logical
device composed of two logical subsystems.We demonstrate that similar
devices can generate two classes of probabilistic propositions :1)
decomposable propositions , which in fact are equivalent to certain ordered
pair of propositions in device subsystems and 2) indecomposable propositions
which are connected with inherent logical interaction between device
units.The indecomposable propositions are undoubtedly of greatest interest
since they , as shown in the paper, provide powerful additional logical
resource compared to standard parallel processing in composite intelligent
systems. The contextual logic of composite devices proposed in this paper,
as we believe, can be also used for analysis of highly organized systems in
cognitive sciences specifically in neuropsychology and linguistics.
\end{abstract}

\maketitle

\section{Introduction}

In the last years close attention of quantum physicists community was
focused on the possibility of creation of effectively working quantum
computer(QC).Although certain progress has been achieved recently in this
field ( see for example \cite{1s} ), nevertheless \ many essential
difficulties yet to be overcome.The most serious obstacle to solve this
problem is undoubtedly decoherence that is interaction of workingr part of
QC with its environment that leads to the destruction of quantum states
superpositions and thus to lower efficiency of quantum computations.Thereby
as we believe it makes sense to pay attention also to other options to
increase computing resourse of intelligent devices which would not be so
sensitive to destruction by inner and exterior noise.One such option
connected with the study of composite probabilistic logical devices just
considered in present paper. The approach proposed uses essentially the
results of preceding authors paper \cite{2s} in which on the basis of
quantum theory of open systems (QTOS) were considered both probabilistic
logical propositions and all possible logical operations (that is logical
connectivies) with them .Thus it becomes possible by simple and unified
method to formulate various logical calculuses: probabilistic logic
(including usual Boolean logic), many- valued logic ( including a modal
logic) and so on.In the present paper our main goal with the help of ideas
and methods of QTOS to state the logic of composite probabilistic devices
that composed of two or more logical units mentally interacting with each
other .Our study can not be considered as too much speculative.Indeed here
is sufficient to point out the well-known fact that brains, both animals and
humans, consist exactly from two hemispheres each of which is able to perfom
special mental functions and continuously exchange with partner by cognitive
information.The rest of the paper is organized as follows.In Section1 for
the convinience of the reader we briefly describe the information from \cite%
{2s} concerning probabilistic logic which is necessary for the understanding
of the given paper.In addition we introduce here the valuable concept of
distanse between two probabilistic propositions with the help of which one
can define the degree of closeness of different propositions.In the Section
2 , that is the main in the paper, we consider the structure of logical
propositions in composite logical devices (CLD).In particular the simplest
example of four -valued logical device is studied more detail.We demonstrate
that all logical propositions in similars system can be divided into two
different classes:1)decomposible propositions (DP) each of which is in fact
equavalent to certain ordered pair of propositions in device logical
subsystems and 2) indecomposible propositions (IP) which are connected with
inherent mental interaction between device subsystems.We believe that the
existence of IP and the possibility to operate with them represents
additional powerful logical resourse in composite systems as compared with
usual parallel information processing.Let us now go to the presentation of
concrete results of the paper.

\section{Preliminary Information}

In this part we briefly remind main results of the paper \cite{2s} which are
necessary for the understanding of the following text.As well as in \cite{2s}
the main subject of our consideration will be a set of plausible
propositions (PP), that is the propositions, the truth or falsity of which
is known to recipient not exactly but only with certain probability.It is
convient to represent any

such proposition by normalized to unity diagonal matrix with positive
elements.For example in the simplest case of two - valued logic any
plausible proposition A can be represented by the matrix :

$\rho \left( A\right) =%
\begin{pmatrix}
p_{A} & 0 \\ 
0 & 1-p_{A}%
\end{pmatrix}%
$, where $p_{A\text{ \ \ }}$is the probability for A to be true.In what
following everywhere where it does not lead to confusion we will identify
propositions with their representative matrices.It turns out that \ various
operations with PP , that is logical connectives, under such approach can be
realized as special transformations of their representative
matrices.Referring the reader for details to $\left( 2\right) $ we represent
here only final decisive result that we will need in the form of the next
theorem.

Theorem1. Let $\rho $ is $N\times N$ diagonal matrix with positive elements
whose trace is equal to 1 and leassume that G is some $M\times N$ matrix
which posseses two defining properties:1) all elements of G are equal 0 or 1
and 2) in each column of G only single element is equal to 1 and all the
rest are equal to zero.In this case the transformation of the form $\ 
\widetilde{\rho }=G\rho G^{T}$ maps $\rho $ onto $M\times M$ diagonal matrix 
$\widetilde{\rho }$ with positive elements $\widetilde{\rho _{ii\text{ }}}$
that satisfy to the relation $\sum\limits_{i=1}^{M}\widetilde{\rho _{ii}}=1$%
.Similar approach can be used in the general case of N- valued probabilistic
( or modal) logic where every proposition has N logical alternatives with
corresponding probabilities $p_{i}$ $\left( i=1,...N\right) $ which satisfy
to normalization condition $\sum\limits_{i=1}^{N}\rho _{i}=1$.We will call
the above transformations as admissible (logical) transformations.Having in
hands the above result one can by unifying way determine all n-place logical
operations in N valued probabilistic logic.For example the result of any
one-place operation applied to PP $A$ can be represented in the form:%
\begin{equation}
\widetilde{A}=G_{1}\cdot A\cdot G_{1}^{T},  \label{a1}
\end{equation}%
where $G_{1}$is some admissible $N\times N$ matrix.

Similarly the result of any two-place logical operation applied to
propositions $B_{1}$ and $B_{2}$ can be written as%
\begin{equation}
\widetilde{B}=G_{2}(B_{1}\otimes B_{2})G_{2}^{T},  \label{a2}
\end{equation}%
where $G_{2}$ is some admissible $N^{2}\times N$ matrix .

It is clear that with the help of admissible transformations one can
determine logical operations for any number of PP as well.Note that in the
case of N-place logical operation it is necessary to take N propositions $%
A_{1},A_{2},...A_{N}$ , and choose as initial state the tensor product of
them ,that is $A_{1}\otimes A_{2}...\otimes A_{N}.$.It is clear also that
total number of one- place logical operations in N -valued logic is equal to 
$N^{N}$, the number of two- place operations is equal to $N^{N^{2}}$ and so
on.Note that the approach proposed also allows one to introduce the concept
of distance between two N valued propositions.Indeed , accoding to natural
reason, the distance $D\left( A,B\right) $ between two propositions $A$ and $%
B$ can be defined as: $D\left( A,B\right) =\frac{1}{2}\sum\limits_{i=1}^{N}%
\mid p_{i}-q_{i}\mid $(we have in mind here that propositions $A$ and $B$
have representative matrices: $diag\{p_{1},p_{i},p_{N}\}$ and $%
diag\{q_{1},q_{i},q_{N}\}$ correspondently.It is clear that above definition
is satisfied to all relevant conditions of the concept.Using this definition
of closeness and also the definitions of basic connectives between
propositions introduced in$\left[ 2\right] $ , one can prove for example the
relation $D\left( A\Longrightarrow B,A\text{ and }B\right) \geqq D\left( B,A%
\text{ and }B\right) $ and many other similar relations.On the other hand
from the physical point of view it is very important that one can think of
representative matrice of proposition as density matrix of relevant quantum
system,and hence should to realize all plausible propositions and logical
operations with them as result of certain physical manupulations in
correspondent quantum system.In addition this analogy allows one to propose
concrete physical realizations of logical devices in order to improve their
effectiveness of information processing.

\section{The logic of composite logical device. The simplest model.}

Till now we consider probabilistic logic and possible logical devices in
which they can be generated as certain integral systems.However, using the
analogy with quantum theory an essential element of which is investigation
of composite systems it is fully justified to include in our focus also
composite logic and correspondently composite logical devices (CLD)
consisting of several subsystems that are logically interconnected with each
other.Evidently such approach is not pure academical.To confirm it enough to
point out that brains of humans and animals exactly consist of two
hemispheres each of which is able to perform special cognitive functions 
\cite{3s}.In this connection it is very essential to emphasize two facts:1)
a flow of afferent information from sensor organs to brain reaches both
hemispheres almost simultaneuosly and 2) left and right hemispheres can
continuously exchange by cognitive information with partner through special
fiber system in the brain, so called the corpus callosum.

Of course the simplest and in many respects primitive model which we will
study in this paper by no means can not represent \ to the right degree any
realistic model of the brain .Nevertheless we believe the very idea about
peculiar logic of CDL deserves close attention and further study.Now let us
turn to the description of composite logic.Note that in this paper in view
of maximal simplicity and clearness we restrict ourselves only the simplest
case of the device in which composite logic should be occur.Namely we will
discuss the four-valued composite logical device consisting of two- valued
subsystems (units).It should be noted however that almost all results of our
consideration can be generalized (with appropriate refinements) to the more
compex situations.Returning to four-valued CDL remind, that according to
Sect.1 any proposition in four- valued logic A can be represented in the
form of diagonal matrix $4\times 4$ , namely: $A=diag%
\{p_{1},p_{2},p_{3},p_{4}\}$ where probabilities of different logical
outcomes $p_{i}$ (i=1..4) satisfy to the conditions:1) $0\leq p_{i}\leq 1$
and 2) $\sum\limits_{i=1}^{4}$ $p_{i}=1$.Our next step is the statement of
basic logical operations (connectives) with such\ propositions.According to
theorem 1 from Sect.1 this task can be realized with the help of relevant
admissible matrices.Omitting intermediate and to some extent tedious
expressions for admissible matrices we adduce here only the required results
for the basic connectives.So , \ the negation of PP A that is $\overline{A}$
can be written as:%
\begin{equation}
\overline{A}=%
\begin{pmatrix}
p_{4} &  &  &  \\ 
& p_{3} &  &  \\ 
&  & p_{2} &  \\ 
&  &  & p_{1}%
\end{pmatrix}%
,  \label{a3}
\end{equation}%
conjuction of two propositions $A$ and $\ B=diag\{q_{1},q_{2},q_{3},q_{4}\}$
is equal to 
\begin{widetext}
\begin{equation}
\left( A\text{ and }B\right) =%
\begin{pmatrix}
p_{1}q_{1} &  &  &  \\ 
& p_{1}q_{2}+p_{2}q_{1}+p_{2}q_{2} &  &  \\ 
&  & p_{1}q_{3}+p_{3}q_{1}+p_{3}q_{3} &  \\ 
&  &  & p_{2}q_{3}+p_{3}q_{2}+p_{4}+q_{4}-p_{4}q_{4}%
\end{pmatrix}%
,  \label{a4}
\end{equation}
\end{widetext}disjunction of the same two propositions A and B is equal to 
\begin{widetext}

\begin{equation}
\left( A\text{ or }B\right) =%
\begin{pmatrix}
p_{3}q_{2}+p_{2}q_{3}+p_{1}+q_{1}-p_{1}q_{1} &  &  &  \\ 
& p_{4}q_{2}+p_{2}q_{4}+p_{2}q_{2} &  &  \\ 
&  & p_{4}q_{3}+p_{3}q_{4}+p_{3}q_{3} &  \\ 
&  &  & p_{4}q_{4}%
\end{pmatrix}%
,  \label{a5}
\end{equation}%

\end{widetext}and so on.Clearly that in the case of four- valued logic there
are considerably more one-place and two- place connectives than in
two-valued case.For example it is easy to see that we have $4^{4}=256$ one
-place connectives in contrast of $2^{2}$ similar connectives in two- valued
logic.Now using the analogy with quantum theory of composite systems \cite%
{4s} we can make the next important step.Let us consider two logical
projections of some four- valued proposition on corresponding two- valued
propositions in each of the subsystem of CLD) .For example arbitrary
proposition $A=diag\{p_{1},p_{2},p_{3},p_{4}\}$ has two logical projections: 
$A_{1}=%
\begin{pmatrix}
p_{1}+p_{2} &  \\ 
& p_{3}+p_{4}%
\end{pmatrix}%
$ in the first subsystem and $A_{2}=%
\begin{pmatrix}
p_{1}+p_{3} &  \\ 
& p_{2}+p_{4}%
\end{pmatrix}%
$ in the second one.On the language of the quantum theory one should say
that the density matrix of composite quantum system uniquely determines
corresponding density matrices of its subsystems.By means of this mapping we
can set the correspondence between all logical operations in composite
system and correspondent operations in its subsystems. It turns out that
above definitions of basic logical connectives in four- valued logic imply
the next simple rules between operations in composite system and
corresponding operations in its subsystems.Indeed one can easily verify that
the simple relations hold:1) $\left( \overline{A}\right) _{1}=\overline{A_{1}%
}$ , $\left( \overline{A}\right) _{2}=\overline{A_{2}}$ \ for negation, and
2) for two -place operations: $\left( A\text{ and }B\right) _{1}=(A_{1}$and $%
B_{1\text{ }})$ , $\left( A\text{ or }B\right) _{1}=(A_{1}$or $B_{1})$.

Now we intend to demonstrate that there are two distinct classes of
propositions generated in composite device, namely 1) decomposable
propositions (DP) each of which in fact is equivalent to some ordered pair
of propositions in device subsystems and 2) indecomposable propositions (IP)
that do not allow such a reduction.To prove this fact it is convinient to
use the representation which is valid in fact for any four- valued
proposition $A$. Namely let \ $A=diag\{p_{1},p_{2},p_{3},p_{4}\}$ then it is
easy to see that $\ A$ can be represented also as 
\begin{widetext}

\begin{equation}
A=%
\begin{pmatrix}
pq+C &  &  &  \\ 
& p\left( 1-q\right) -C &  &  \\ 
&  & q\left( 1-p\right) -C &  \\ 
&  &  & \left( 1-p\right) \left( 1-q\right) +C%
\end{pmatrix}%
,  \label{a6}
\end{equation}

\end{widetext}where in Eq. (\ref{a6}) we use the notation: $\ \
p=p_{1}+p_{2},$ $\ q=p_{1}+p_{3}$ and $C=p_{1}p_{4}-p_{2}p_{3}.$

The collection of propositions which satisfy to condition $C=0$ \ will be
called decomposable since for them the decomposition in the form of tensor
product :$A=A_{1}\otimes A_{2}$ obviously holds.Thus such "composite"
propositions are in fact coincided with tensor products of their logical
projections.Note that in the quantum theory of composite systems analogues
of these propositions are factorable pure states and correspondent quantity $%
C$ (that is equal to zero in the case of factorable states) is called
concurrence.However,with respect to the case of probabilistic many-valued
logic, will call the quantity $C$ from Eq. (\ref{a6}) the context variable
or simply as context.Reasons for such designation will become clear a little
later.Now let us point out two nearly obvious facts concerning DP:1) if $A$
is DP then $\overline{A}$ is DP as well and 2) if $A$ and $B$ are two DP
then propositions ($A$ and $B$) and ($A$ or $B$) are decomposable
also.Thereby the collection of all DP is closed with respect to basic
logical transformations, namely one can prove that the following theorem has
place.Theorem 2 : let $A$ is DP in composite system and $G$ some one-place
admissible transformation.Then $\widetilde{A}=GAG^{T}$ is decomposable
proposition as well and can uniquely \ be represented in the form: $%
\widetilde{A}=\widetilde{A_{1}}\otimes \widetilde{A_{2}}$ where $\widetilde{%
A_{1}}=G_{1}\left( A_{1}\otimes A_{2}\right) G_{1}^{T}$ \ and $\widetilde{%
A_{2}}=G_{2}(A_{1}\otimes A_{2})G_{2}^{T}$ where $\left( G_{1},G_{2}\right) $
is some ordered pair of two-place admissible transformations and $A_{1},A_{2%
\text{\ }}$are logical projections of DP $A$ in device subsystems.Without
stopping to prove theorem 2 note only that total number of admissible
one-place transformations in CLD is equal $4^{4}$ which is coincides with
the number of ordered two- place pairs $\left( G_{1,}G_{2}\right) $ which is
equal to $2^{4}\times 2^{4}.$

Thus we come to the important conclusion: all logical operations with DP in
CLD can be entirely reduced to the logical operations produced in device
subsystems that is in this case CLD works exactly according to principles of
parallel processing information.On the other hand we argue that for
indecomposable propositions the situation is quite different.It turns out
that existence of IP and the possibility to operate with them to a large
degree increase logical resourses of CLD.This important topic undoubted
deserves a special and detail investigation. But in this paper which has
preliminary nature we restrict ourselves only to single but very good
example illustrating the above thesis.Let us consider in CLD one parameter
collection of indecomposable propositions of the next form:%
\begin{equation}
A\left( C\right) =%
\begin{pmatrix}
\frac{1}{4}+C &  &  &  \\ 
& \frac{1}{4}-C &  &  \\ 
&  & \frac{1}{4}-C &  \\ 
&  &  & \frac{1}{4}+C%
\end{pmatrix}%
,  \label{a7}
\end{equation}%
where the variable $C$ satisfies to unequalities:$\left( -\frac{1}{4}\leq
C\leq \frac{1}{4}\right) .$ Evidently that if $C\neq 0$ then all
propositions $A\left( C\right) $ are indecomposable and the context variable
for proposition $A\left( C\right) $ coincides exactly with $C.$Note in
passing that for any IP $A$ its distance to the nearest DP that is $%
A_{1}\otimes A_{2}$ is equal exactly to 4$\left\vert C\right\vert $ where $C$
is the value of context variable for $A.$It is easy to see also that logical
projections for proposition $A\left( C\right) $ from $\left( 7\right) $ are
independent from context $C$ and equal to each other, namely $A_{1}=A_{2}=%
\begin{pmatrix}
\frac{1}{2} &  \\ 
& \frac{1}{2}%
\end{pmatrix}%
.$Now let us perform the concrete one-place logical operation :$\widetilde{A}%
\left( C\right) =GA\left( C\right) G^{T}$, where the admissible matrix $G$
has the form:%
\begin{equation}
G=%
\begin{pmatrix}
0 & 1 & 0 & 0 \\ 
1 & 0 & 0 & 0 \\ 
0 & 0 & 1 & 0 \\ 
0 & 0 & 0 & 1%
\end{pmatrix}%
,  \label{a8}
\end{equation}%
Note that above operation is isometric that is it does not change the
distances between propositions from collection of interest .Really, it is
easy to verify that: $D\left[ A\left( C_{1}\right) ,A\left( C_{2}\right) %
\right] =D\left[ \widetilde{A\left( C_{1}\right) },\widetilde{A\left(
C_{2}\right) }\right] =4\left\vert C_{1}-C_{2}\right\vert .$ In explicit
form the matrix $\widetilde{A\left( C\right) }$ reads as%
\begin{equation}
\widetilde{A\left( C\right) }=%
\begin{pmatrix}
\frac{1}{4}-C &  &  &  \\ 
& \frac{1}{4}+C &  &  \\ 
&  & \frac{1}{4}-C &  \\ 
&  &  & \frac{1}{4}+C%
\end{pmatrix}%
,  \label{a9}
\end{equation}%
The expression Eq. (\ref{a9}) implies that logical projections of $%
\widetilde{A\left( C\right) }$ \ in device subsystems has the form: $%
\widetilde{A_{1}}=%
\begin{pmatrix}
\frac{1}{2} &  \\ 
& \frac{1}{2}%
\end{pmatrix}%
$ and $\widetilde{A_{2}}=%
\begin{pmatrix}
\frac{1}{2}-2C &  \\ 
& \frac{1}{2}+2C%
\end{pmatrix}%
$. We see that logical projection of transformed proposition in the first
subsystem does not change, while its projection in the second subsystem
essentially depends from context value $C$.Moreover the very plausibility of
proposition $\widetilde{A_{2}}$ may be determined by context and depending
on $C$ can change its value to the contrary.Indeed, the value of $C=-\frac{1%
}{4}$ implies that $\widetilde{A_{2}}\left( -\frac{1}{4}\right) =%
\begin{pmatrix}
1 &  \\ 
& 0%
\end{pmatrix}%
$and value of $C=\frac{1}{4}$ implies that $\ \widetilde{A_{2}}\left( \frac{1%
}{4}\right) =%
\begin{pmatrix}
0 &  \\ 
& 1%
\end{pmatrix}%
$.In this connection one can say that \ the applied logical operation is
connected with recognition of context by second subsystem (agent).Of course
the similar transformation can be applied with respect to first subsystem
(agent) as well.We are convinced that application of IP and logical
operations with them allows one significantly expend the class of deduced
propositions in subsystems of CLD comparing with ordinary Boolean functions
from $p$ and $q$.that are realized in the case of decomposable propositions.

Thus already this preliminary analysis of simple CLD clealy demonstrates the
advantages of such devices in logical processing in comparance with ordinary
parallel processing systems.

In conclusion we emphazise \ that many important and interesting issues have
remained outside of our scope .Among the most important topics we may call
such as : 1) the need to generalize the approach proposed on the case of
composite system consisting of arbitrary number of subsystems with different
dimensions 2) the need to classify logical operations in CDL from the point
of view of their possibility to transform DP into IP and vica versa.Besides
it is necessary to study in detail the problem of realization CDL in
concrete quantum physical systems.In particular in the case of simplest
four-valued CDL we should talk about two-qubit quantum system interacting
with environment by special way.In part we consider this issue in \cite{2s},
but this question needs much more detail investigation.

All these and some other questions we hope to discuss in later publications..


\begin{thebibliography}{9}
\bibitem{1s} J.J. Pla, K.Y. Tan, J.P. Dehollain, W.H. Lim, J.J. Morton, D.N.
Jamieson, A.S.Dzurak, A.Morello, Nature, 489, (7417), 541, (2012).

\bibitem{2s} E. D. Vol, Int. J. Theor. Phys, 52, 514, (2013).

\bibitem{3s} S. P. Springer, Left Brain, Right Brain, W. H. Freeman, New
York, (1989).

\bibitem{4s} M. A. Nielsen and I. L. Chuang, Quantum Computation and Quantum
Information, Cambridge University Press, Cambridge,(2001).
\end{thebibliography}
\end{document}